\documentclass{ws-procs9x6}            % default, citations in superscript
\begin{document}
\title{Update on Dusty Sources and Candidate Young Stellar Objects in the S-cluster}

\author{Maria Melamed$^*$ and Florian Pei$\beta$ker}

\address{I.Physikalisches Institut der Universit\"at zu K\"oln, \\ Z\"ulpicher Str. 77, 50937 K\"oln, Germany\\
$^*$E-mail: melamed@ph1.uni-koeln.de}

%\author{A. N. Author}

%\address{Group, Laboratory, Street,\\
%City, State ZIP/Zone, Country\\
%E-mail: an\_author@laboratory.com}

\begin{abstract}
The Galactic Center provides a unique opportunity to observe a galactic core, objects in close proximity to a supermassive black hole (SMBH), and star formation channels that exhibit imprints of this peculiar environment. This habitat hosts, in addition to the SMBH Sgr A*, a surprisingly young cluster with the so-called S-stars. These stars orbit the SMBH on timescales of a few years with thousands of km/s. While the presence of high-velocity stars in the S-cluster already raises a variety of scientific questions, the observation of several bright L-band emission sources has resulted in a rich discussion of their nature. The detection of a prominent Doppler-shifted Br$\gamma$ line accompanies most of these sources that seem to be embedded in a dusty envelope. Using the radiative-transfer model HYPERION, we find strong indications of the presence of a stellar low-mass population embedded in the S-cluster. We revisit this intriguing cluster and its dusty members that orbit the supermassive black hole Sgr A* on bound Keplerian trajectories. Among these cluster members, there is one source that initiated the studies of this analysis: G1. We find that the flux density of G1 in the NIR and MIR resembles a spectral energy distribution of
a Class I YSO, which contributes to the “Paradox of Youth”.
\end{abstract}


\bodymatter

\section{Introduction}\label{aba:secIntro}
The central region of the Milky Way, known as the Galactic Center (GC), allows us to analyze a highly dynamical part of our galaxy of significant scientific interest in a unique way and thus is an object of extensive research. It is now well established from a variety of studies, that the GC hosts the supermassive black hole (SMBH) Sagitarrius A* (Sgr A*) \cite{Eckart2017}. Thus, the analysis of the GC does not only allow for detailed studies of such an object, but also objects close to SMBH which we could not yet resolve in other galaxies. In recent years, there has been extensive research on potential star formation channels in this region \cite{Morris1996, Eckart2004, Genzel2010, Schoedel2011S, Pfuhl2011, Schodel2020, Peissker2023X3, Peissker2024IRS13II}. This was particularly motivated by the "Paradox of Youth", which describes the surprising finding of young OB-type stars known as the S cluster stars (S stars) in the inner parsec of the GC \cite{Ghez2003, Genzel2010}. These S stars orbit Sgr A* at very high velocities on timescales of a few to tens of years in most cases. \\
Within the S cluster, the observation of several other sources with high L-band luminosity and prominent emission of lines such as the Brackett-$\gamma$ line raised the interest in the GC. The nature of these objects, known as G objects or dusty (D) sources, is still under debate. For example, some studies claim that they must be pure gas clouds, while others postulate a stellar component \cite{Gillessen2012, Burkert2012, MurrayClay2012, Eckart2013NIRProperMotions, Jalali2014, Zajacek2014, Valencia2015, Plewa2017, Ciurlo2020Natur, Peissker2021G2}.\\
This paper aims to summarize some of the most recent publications on these sources and examines the relationship of their results.

\section{Temperature Mapping}
\begin{figure}
\begin{center}
\includegraphics[width=4.5in]{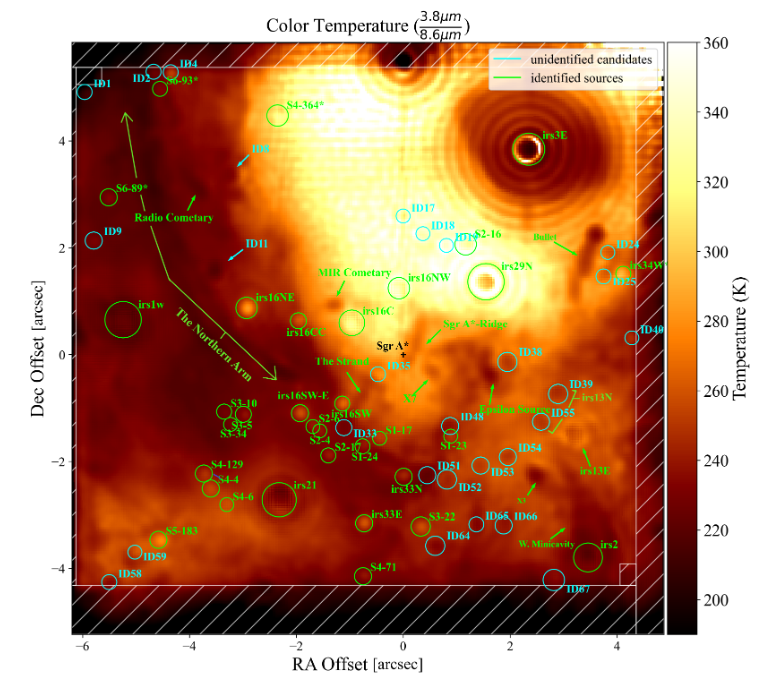}
\end{center}
\caption{Color temperature map of the Galactic Center. Compact and extended features marked in green if known from former publications, and in cyan if identified newly \cite{Dinh2024}.}%"Color temperature map, annotated with all compact objects and extended features. Previously known features are marked in green, newly reported ones are marked in cyan."\cite{Dinh2024}}
\label{aba:Dinh3}
\end{figure}

\begin{figure}
\begin{center}
\includegraphics[width=4.5in]{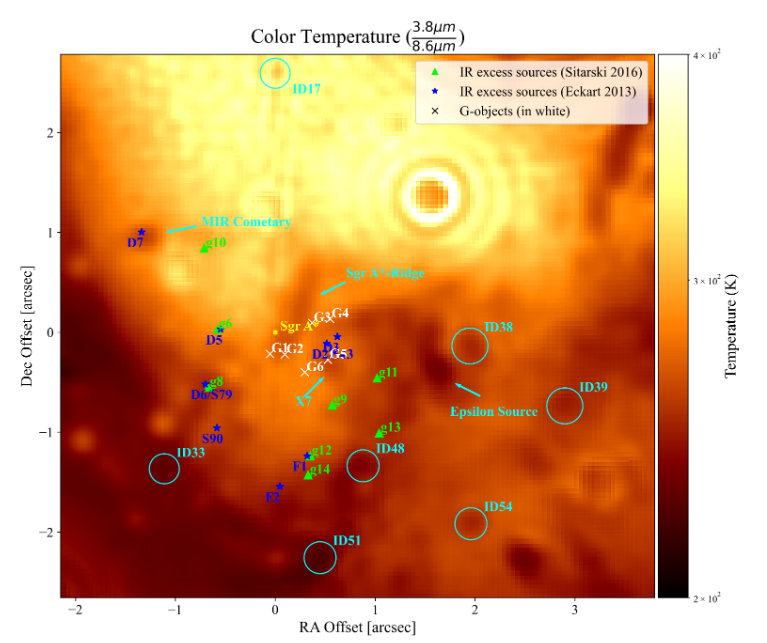}
\end{center}
\caption{Color temperature map of the Galactic Center marking the locations of G objects, infrared excess sources, and cold sources in the vicinity. Several cool sources are located outside of the portrayed area. Sgr A*-ridge might be more distant then suggested by image as it has a different position in the line of sight.\cite{Dinh2024}}%"Locations of currently known Infrared Excess sources and G objects, compared to the identified cold sources within the vicinity. Cool sources ID5 4, 55, and 65 are located beyond this region. The so-called Sgr A*-ridge superimposed nearby Sgr A*on the line of sight and not actually within its vicinity in 3D (but see also Peißker et al. 2020 \cite{Peissker2020IRObserv})." \cite{Dinh2024}}
\label{aba:Dinh5}
\end{figure}

While several studies on G objects have already been conducted, it is plausible that not all G objects have been identified yet. Dinh et al. (2024) \cite{Dinh2024} analyzed mid-infrared NIRC2/Keck and VISIR/VLT data. From this data, a color-temperature map is obtained (see Figure \ref{aba:Dinh3}), where compact and point-like features stand out. This features are classified as "cold", when they show temperatures of 10\,K colder than their background, and "hot" when they show temperatures of 10\,K hotter than their background. While 22 "hot" sources in the temperature map are associated with known stars, 11 "cold" sources are newly identified. As known G objects are also cold compact sources, these newly identified sources share this property and are included as candidate G objects in Figure \ref{aba:Dinh5}. \\
It should be noted that the possible membership to the G population also depends on the exact definition of this G population: While different publications agree that G objects are compact sources that show Brackett-$\gamma$ line emission, and most publications identify them in near-infrared continuum emission, statements regarding the extent of the population vary. Thus, further analysis to confirm these properties is necessary to determine whether the suggested sources are indeed G objects. A consensus on the exact definition and requirements of this population will be beneficial for the classification.

%version davor:
%Dinh et al. (2024) \cite{Dinh2024} identified 11 new objects that might be G sources, and present a comprehensive list of candidate G objects. These results are based on mid-infrared NIRC2/Keck and VISIR/VLT data. From this data, a color-temperature map is obtained (see Figure \ref{aba:Dinh3}), where previously known sources stand out, along with several newly identified sources. The new candidate G objects are identified (Figure \ref{aba:Dinh5}) by comparing this map to previous results and known sources. %Comparing this map to previous results and known sources, the new candidate G objects are identified (Figure \ref{aba:Dinh5}). %xxx...

\section{Kinematic Structure} \label{aba:secKinematicStructure}
\begin{figure}
\begin{center}
\includegraphics[width=4.5in]{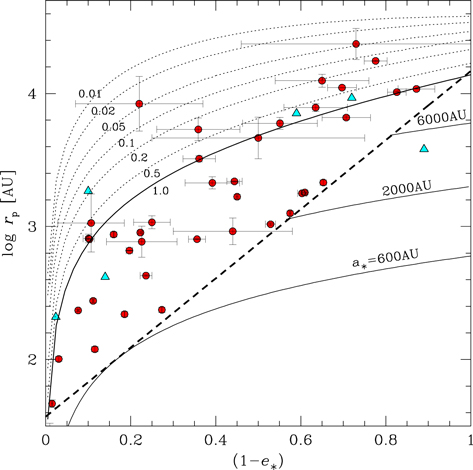}
\end{center}
\caption{Plot of pericenter distance depending on orbital eccentricity. Red points mark S stars, cyan triangles mark G sources. Dotted lines with labeled constant probability represent how likely the orbits of the S stars are determined in 20 years of observational data. The solid line shows 1.0 probability, thus all orbital parameters of stars should be possible to find. Dashed line represents the boundary below which the "Zone of Avoidance" is located. \cite{Burkert2024}}%"Dotted lines show contours of constant probability P (see labels) to determine the orbits of S-stars within 20 yr of observations. It should be possible to determine the orbital parameters for all stars with pericenter distances below the solid black line, labeled 1.0. Red points with error bars show those S-stars for which orbital parameters have been determined. Cyan triangles show G-clouds. The dashed black line marks the upper boundary of the zone of avoidance within which no stars with known orbital parameters are found, even though it should be easy to determine their orbital parameters. For reference, the solid lines in the zone of avoidance show positions of constant semimajor axis a."\cite{Burkert2024}}
\label{aba:Burkert3}
\end{figure}

Studies of the dynamics of the S stars report a distribution of the stars where they are arranged in discs \cite{Ali2020}. Burkert et al. (2024) \cite{Burkert2024} revisited these studies and investigated the orbital parameters and their relations. The authors find a strong dependence of the pericenter distance $r_p$ and orbital eccentricity $e$ which is non-random. Moreover, as visible in Figure \ref{aba:Burkert3}, no sources with high $r_p$ and $e$ are found. Qualitative and stochastic analysis show that this is not by chance or lack of data, but appears to indeed be an unpopulated region. The authors call this region the "Zone of Avoidance". This finding puts further constraints on possible formation scenarios of the cluster.\\
In addition to the S stars, this study takes into account the orbits of the G objects. Interestingly, the G objects are found in the same distribution as the S stars. They also do not fall into the "Zone of Avoidance". Although the authors stress the limitations of the study from the small number of considered G objects, this result is relevant to understand the possible relationship between the S stars and G objects, as it reveals additional commonalities.\\

\begin{figure}
\begin{center}
\includegraphics[width=0.32\textwidth]{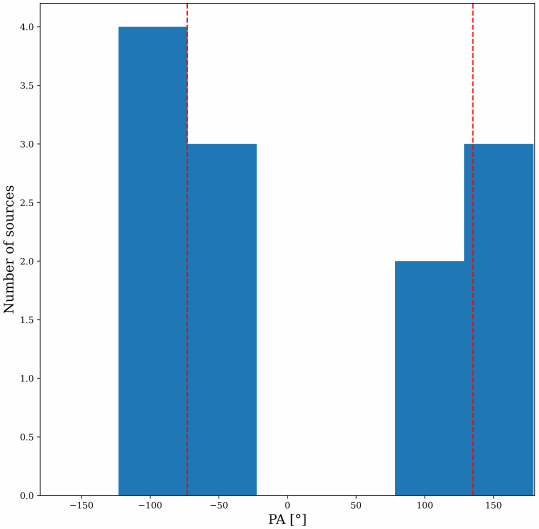}
\includegraphics[width=0.32\textwidth]{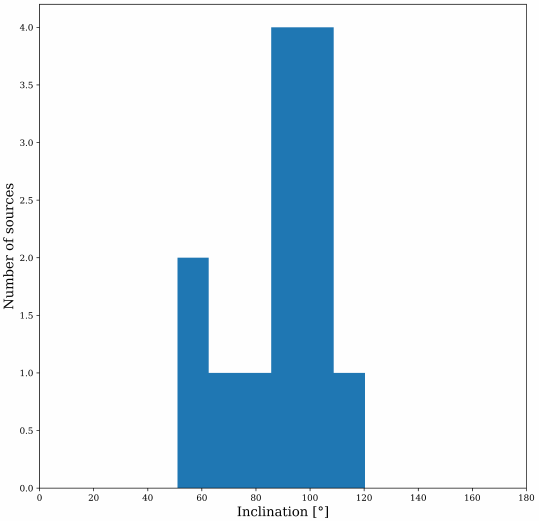}
\includegraphics[width=0.32\textwidth]{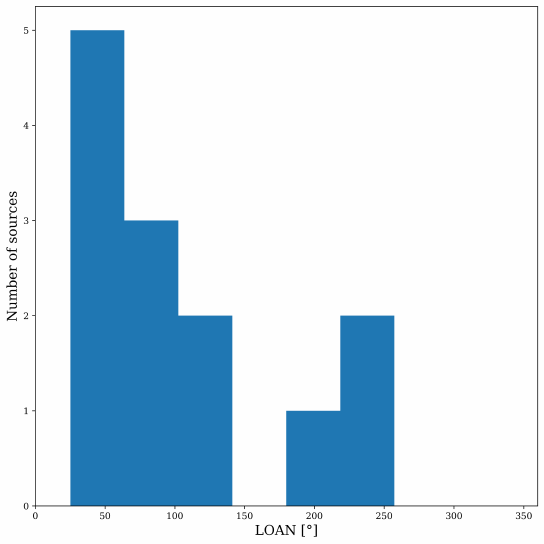}

\end{center}
\caption{From left to right: Position angle, inclination distribution, and distribution of the longitudes of the ascending node (LOAN) for the dusty sources \cite{Peissker2024YSOScluster}. The position angle and LOAN show a non-randomized and bimodal distribution. The inclination shows a strong correlation to the "red disk" of the S stars \cite{Ali2020, Peissker2024YSOScluster}.}
\label{aba:Peissker5-7}
\end{figure}

These results are further confirmed by Peissker et al. (2024)\cite{Peissker2024YSOScluster}, where more observations and sources are considered, and new orbits for several dusty objects are determined. The analysis of the position angles, longitudes of the ascending node, and inclinations (see Figure \ref{aba:Peissker5-7}) reveal a distribution that is similar to the S stars and arranged in one disk: This disk coincides with the "red" disk proposed in Ali et al. (2020)\cite{Ali2020}. These similarities support a closer connection and possible commonalities in the origin between the S stars and G objects. \\

\section{Nature of the G Objects}
\begin{figure}
\begin{center}
\includegraphics[width=4.5in]{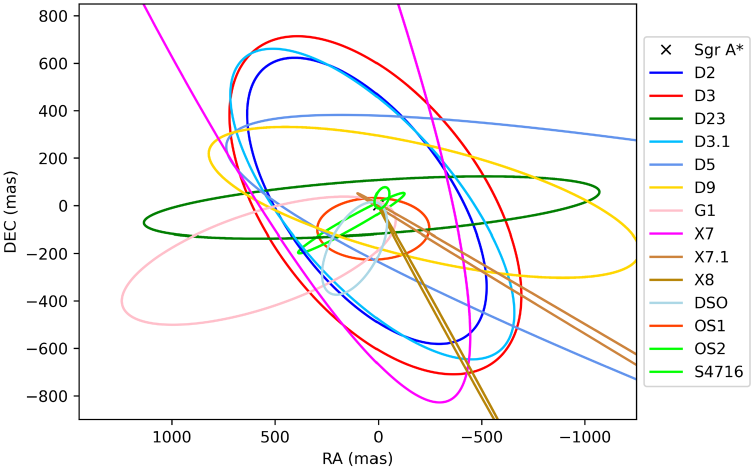}
\end{center}
\caption{Overview of the Keplarian orbits of the analyzed dusty objects. This Keplerian orbits result from the best fit solutions. Results are limited by time frame of observational data \cite{Peissker2024YSOScluster}}%"Keplerian orbits of all the investigated dusty objects in this work. The shown Keplerian solutions represent best fit orbits that are limited by the data baseline. Therefore, more data results in a longer data baseline with increased accuracy. Here Sgr A* is located at (0,0). The orbit of S4716 is taken from \cite{Peissker2022}. North is up and east is to the left."\cite{Peissker2024YSOScluster}}%xxx
\label{aba:Peissker4}
\end{figure}

\begin{figure}
\begin{center}
\includegraphics[width=4.5in]{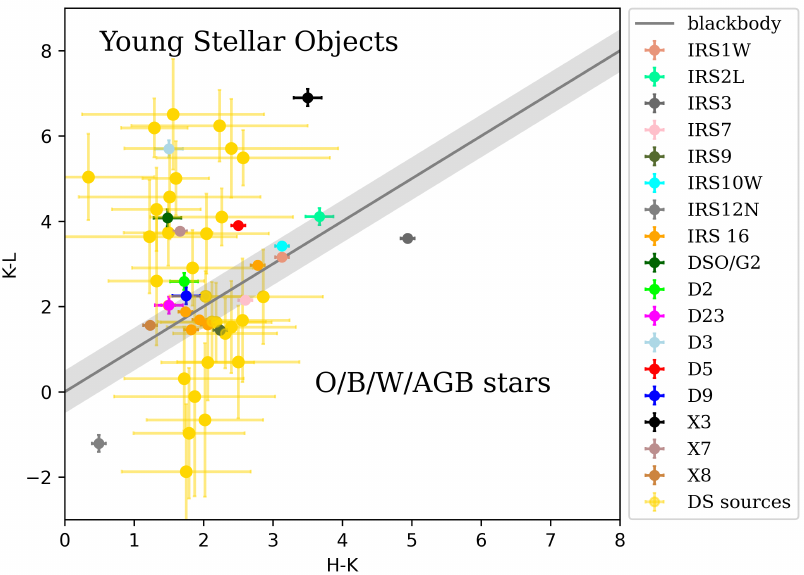}
\end{center}
\caption{Color-color diagram showing dusty sources in the Galactic Center. Dusty sources located within the S cluster marked in different colors as shown in the legend, dusty sources (DS) in IRS13 marked in gold.\cite{Peissker2024YSOScluster}} % "Color–color diagram of the dusty source of the S-cluster and IRS13 (in gold). Although the IRS13 sample is slightly larger compared to the S-cluster objects investigated in this work, both groups of dusty sources show similarities, implying a shared nature. Especially the dusty sources of the S-cluster seem to share a comparable photometric footprint. All uncertainties reflect the standard deviation of the related magnitude estimated for all available epochs between 2006 and 2019. The linear gray fit represents a one-component blackbody. Based on this classification, we conclude that the majority of dusty sources exhibit colors that are consistent with a YSO classification. As a comparison, we include the H–K and K–L colors for several main-sequence and AGB stars from Blum et al. (1996), Ott et al. (1999), Pott et al. (2008), and Peißker et al. (2023b)."\cite{Peissker2024YSOScluster}} %xxx
\label{aba:Peissker8}
\end{figure}

The nature of the G objects has been the subject of a lot of studies as it provides valuable insights into understanding crucial aspects of the GC, such as possible star formation channels. The similarities between the S stars and G objects discussed in the previous section already provide relevant constraints to this problem. Peissker et al. (2024) \cite{Peissker2024YSOScluster} analyze data from SINFONI/VLT, NACO/VLT, and NIRC2/Keck from observations between 2005 and 2019 to investigate the dusty sources G1, G2/DSO, D2/G3, D23, D3, D3.1, D5, D9, OS1, OS2, X7, X7.1/G5, and X8. The sources are identified in the various data sets for a multiwavelength dynamical and photometric analysis. \\
It should be noted that the dynamical analysis does not only show that the G sources exist within the "red" disk, but that Keplerian orbits for the considered objects can be determined (see Figure \ref{aba:Peissker4}) \cite{Peissker2024YSOScluster}. The identification of the dusty sources across the analyzed wavebands, and their movement as compact sources on Keplerian orbits make a purely dusty nature unlikely and indicates the presence of a stellar component. The similarities between the S stars and G objects could be explained by either a common formation scenario or a common channel in formation and migration. \\
Another relevant result of the publication is the color-color diagram of the G objects (Figure \ref{aba:Peissker8}).  The G objects show colors comparable to those of young stellar objects (YSOs). Thus, this possible YSO nature is further investigated by modelling the spectral energy distributions (SEDs) for the individual sources using the code HYPERION \cite{Robitaille2011}. The photometric data points for the dusty sources can indeed be described by Class I YSO models in all cases except for the source D9. \\

A more comprehensive analysis of G1 will be provided in Melamed et al. (in prep.). In this forthcoming analysis, the previous identification of G1 in the L' band will be complemented by its identification in the K band. As the K band and the ionized hydrogen correspond to temperatures that exceed a coreless gas cloud to survive throughout the epochs of observation \cite{Peissker2024YSOScluster}, this result strongly supports the model of G1 with a stellar component \cite{Shahzamanian2016, Witzel2017, Zajacek2017}. Considering the photometric analysis, G1 as well as the other dusty sources can be interpreted as candidate YSOs.\\
These results provide further constraints on possible star formation scenarios in the GC and thus offer valuable insights to a better understanding the Paradox of Youth.

%xxx add acknowledgements!!!
\section*{Acknowledgements}
We gratefully acknowledges the Collaborative Research Center 1601 funded by the Deutsche Forschungsgemeinschaft (DFG, German Research Foundation) – SFB 1601 [sub-project A3] - 500700252. MM is a member of the International Max Planck Research School for Astronomy and Astrophysics at the Universities of Bonn and Cologne (IMPRS) and the Bonn Cologne Graduate School (BCGS), and appreciates their support.  %BCGS as well?xxx

\bibliographystyle{ws-procs9x6} % for numbered citation & references
\bibliography{ws-procs9x6}

\begin{thebibliography}{10}

\bibitem{Eckart2017}
A.~{Eckart}, A.~{H{\"u}ttemann}, C.~{Kiefer}, S.~{Britzen}, M.~{Zaja{\v{c}}ek}, C.~{L{\"a}mmerzahl}, M.~{St{\"o}ckler}, M.~{Valencia-S}, V.~{Karas} and M.~{Garc{\'\i}a-Mar{\'\i}n}, {The Milky Way's Supermassive Black Hole: How Good a Case Is It?}, {\em Foundations of Physics} {\bf 47}, 553 (May 2017).

\bibitem{Morris1996}
M.~{Morris} and E.~{Serabyn}, {The Galactic Center Environment}, {\em Annual Review of Astron and Astrophys} {\bf 34}, 645 (January 1996).

\bibitem{Eckart2004}
A.~{Eckart}, J.~{Moultaka}, T.~{Viehmann}, C.~{Straubmeier} and N.~{Mouawad}, {Young Stars at the Center of the Milky Way?}, {\em Astrophysical Journal} {\bf 602}, 760 (February 2004).

\bibitem{Genzel2010}
R.~{Genzel}, F.~{Eisenhauer} and S.~{Gillessen}, {The Galactic Center massive black hole and nuclear star cluster}, {\em Reviews of Modern Physics} {\bf 82}, 3121 (Oct 2010).

\bibitem{Schoedel2011S}
R.~{Sch{\"o}del}, {The center of the Milky Way} (November 2011).

\bibitem{Pfuhl2011}
O.~{Pfuhl}, T.~K. {Fritz}, M.~{Zilka}, H.~{Maness}, F.~{Eisenhauer}, R.~{Genzel}, S.~{Gillessen}, T.~{Ott}, K.~{Dodds-Eden} and A.~{Sternberg}, {The Star Formation History of the Milky Way's Nuclear Star Cluster}, {\em Astrophysical Journal} {\bf 741}, p. 108 (November 2011).

\bibitem{Schodel2020}
R.~{Sch{\"o}del}, F.~{Nogueras-Lara}, E.~{Gallego-Cano}, B.~{Shahzamanian}, A.~T. {Gallego-Calvente} and A.~{Gardini}, {The Milky Way's nuclear star cluster: Old, metal-rich, and cuspy. Structure and star formation history from deep imaging}, {\em Astronomy and Astrophysics} {\bf 641}, p. A102 (September 2020).

\bibitem{Peissker2023X3}
F.~{Pei{\ss}ker}, M.~{Zaja{\v{c}}ek}, N.~B. {Sabha}, M.~{Tsuboi}, J.~{Moultaka}, L.~{Labadie}, A.~{Eckart}, V.~{Karas}, L.~{Steiniger}, M.~{Subroweit}, A.~{Suresh}, M.~{Melamed} and Y.~{Cl{\'e}net}, {X3: A High-mass Young Stellar Object Close to the Supermassive Black Hole Sgr A*}, {\em Astrophysical Journal} {\bf 944}, p. 231 (February 2023).

\bibitem{Peissker2024IRS13II}
F.~{Pei{\ss}ker}, M.~{Zaja{\v{c}}ek}, M.~{Labaj}, L.~{Thomkins}, A.~{Elbe}, A.~{Eckart}, L.~{Labadie}, V.~{Karas}, N.~B. {Sabha}, L.~{Steiniger} and M.~{Melamed}, {The Evaporating Massive Embedded Stellar Cluster IRS 13 Close to Sgr A*. II. Kinematic Structure}, {\em Astrophysical Journal} {\bf 970}, p.~74 (July 2024).

\bibitem{Ghez2003}
A.~M. {Ghez}, G.~{Duch{\^e}ne}, K.~{Matthews}, S.~D. {Hornstein}, A.~{Tanner}, J.~{Larkin}, M.~{Morris}, E.~E. {Becklin}, S.~{Salim}, T.~{Kremenek}, D.~{Thompson}, B.~T. {Soifer}, G.~{Neugebauer} and I.~{McLean}, {The First Measurement of Spectral Lines in a Short-Period Star Bound to the Galaxy's Central Black Hole: A Paradox of Youth}, {\em Astrophysical Journal, Letters} {\bf 586}, L127 (April 2003).

\bibitem{Gillessen2012}
S.~{Gillessen}, R.~{Genzel}, T.~K. {Fritz}, E.~{Quataert}, C.~{Alig}, A.~{Burkert}, J.~{Cuadra}, F.~{Eisenhauer}, O.~{Pfuhl}, K.~{Dodds-Eden}, C.~F. {Gammie} and T.~{Ott}, {A gas cloud on its way towards the supermassive black hole at the Galactic Centre}, {\em Nature} {\bf 481}, 51 (January 2012).

\bibitem{Burkert2012}
A.~{Burkert}, M.~{Schartmann}, C.~{Alig}, S.~{Gillessen}, R.~{Genzel}, T.~K. {Fritz} and F.~{Eisenhauer}, {Physics of the Galactic Center Cloud G2, on Its Way toward the Supermassive Black Hole}, {\em Astrophysical Journal} {\bf 750}, p.~58 (May 2012).

\bibitem{MurrayClay2012}
R.~A. {Murray-Clay} and A.~{Loeb}, {Disruption of a proto-planetary disc by the black hole at the milky way centre}, {\em Nature Communications} {\bf 3}, p. 1049 (September 2012).

\bibitem{Eckart2013NIRProperMotions}
A.~{Eckart}, K.~{Mu{\v{z}}i{\'c}}, S.~{Yazici}, N.~{Sabha}, B.~{Shahzamanian}, G.~{Witzel}, L.~{Moser}, M.~{Garcia-Marin}, M.~{Valencia-S.}, B.~{Jalali}, M.~{Bremer}, C.~{Straubmeier}, C.~{Rauch}, R.~{Buchholz}, D.~{Kunneriath} and J.~{Moultaka}, {Near-infrared proper motions and spectroscopy of infrared excess sources at the Galactic center}, {\em Astronomy and Astrophysics} {\bf 551}, p. A18 (March 2013).

\bibitem{Jalali2014}
B.~{Jalali}, F.~I. {Pelupessy}, A.~{Eckart}, S.~{Portegies Zwart}, N.~{Sabha}, A.~{Borkar}, J.~{Moultaka}, K.~{Mu{\v{z}}i{\'c}} and L.~{Moser}, {Star formation in the vicinity of nuclear black holes: young stellar objects close to Sgr A*}, {\em Monthly Notices of the RAS} {\bf 444}, 1205 (October 2014).

\bibitem{Zajacek2014}
M.~{Zaja{\v{c}}ek}, V.~{Karas} and A.~{Eckart}, {Dust-enshrouded star near supermassive black hole: predictions for high-eccentricity passages near low-luminosity galactic nuclei}, {\em Astronomy and Astrophysics} {\bf 565}, p. A17 (May 2014).

\bibitem{Valencia2015}
M.~{Valencia-S.}, A.~{Eckart}, M.~{Zaja{\v{c}}ek}, F.~{Peissker}, M.~{Parsa}, N.~{Grosso}, E.~{Mossoux}, D.~{Porquet}, B.~{Jalali}, V.~{Karas}, S.~{Yazici}, B.~{Shahzamanian}, N.~{Sabha}, R.~{Saalfeld}, S.~{Smajic}, R.~{Grellmann}, L.~{Moser}, M.~{Horrobin}, A.~{Borkar}, M.~{Garc{\'\i}a-Mar{\'\i}n}, M.~{Dov{\v{c}}iak}, D.~{Kunneriath}, G.~D. {Karssen}, M.~{Bursa}, C.~{Straubmeier} and H.~{Bushouse}, {Monitoring the Dusty S-cluster Object (DSO/G2) on its Orbit toward the Galactic Center Black Hole}, {\em Astrophysical Journal} {\bf 800}, p. 125 (February 2015).

\bibitem{Plewa2017}
P.~M. {Plewa}, S.~{Gillessen}, O.~{Pfuhl}, F.~{Eisenhauer}, R.~{Genzel}, A.~{Burkert}, J.~{Dexter}, M.~{Habibi}, E.~{George}, T.~{Ott}, I.~{Waisberg} and S.~{von Fellenberg}, {The Post-pericenter Evolution of the Galactic Center Source G2}, {\em Astrophysical Journal} {\bf 840}, p.~50 (May 2017).

\bibitem{Ciurlo2020Natur}
A.~{Ciurlo}, R.~D. {Campbell}, M.~R. {Morris}, T.~{Do}, A.~M. {Ghez}, A.~{Hees}, B.~N. {Sitarski}, K.~{Kosmo O'Neil}, D.~S. {Chu}, G.~D. {Martinez}, S.~{Naoz} and A.~P. {Stephan}, {A population of dust-enshrouded objects orbiting the Galactic black hole}, {\em Nature} {\bf 577}, 337 (January 2020).

\bibitem{Peissker2021G2}
F.~{Pei{\ss}ker}, M.~{Zaja{\v{c}}ek}, A.~{Eckart}, B.~{Ali}, V.~{Karas}, N.~B. {Sabha}, R.~{Grellmann}, L.~{Labadie} and B.~{Shahzamanian}, {The Apparent Tail of the Galactic Center Object G2/DSO}, {\em Astrophysical Journal} {\bf 923}, p.~69 (December 2021).

\bibitem{Dinh2024}
C.~K. {Dinh}, A.~{Ciurlo}, M.~R. {Morris}, R.~{Sch{\"o}del}, A.~{Ghez} and T.~{Do}, {High-resolution, Mid-infrared Color Temperature Mapping of the Central 10\" of the Galaxy}, {\em Astronomical Journal} {\bf 167}, p.~41 (January 2024).

\bibitem{Ali2020}
B.~{Ali}, D.~{Paul}, A.~{Eckart}, M.~{Parsa}, M.~{Zajacek}, F.~{Pei{\ss}ker}, M.~{Subroweit}, M.~{Valencia-S.}, L.~{Thomkins} and G.~{Witzel}, {Kinematic Structure of the Galactic Center S Cluster}, {\em Astrophysical Journal} {\bf 896}, p. 100 (June 2020).

\bibitem{Burkert2024}
A.~{Burkert}, S.~{Gillessen}, D.~N.~C. {Lin}, X.~{Zheng}, P.~{Schoeller}, F.~{Eisenhauer} and R.~{Genzel}, {The Orbital Structure and Selection Effects of the Galactic Center S-star Cluster}, {\em Astrophysical Journal} {\bf 962}, p.~81 (February 2024).

\bibitem{Peissker2024YSOScluster}
F.~{Pei{\ss}ker}, M.~{Zaja{\v{c}}ek}, M.~{Melamed}, B.~{Ali}, M.~{Singhal}, T.~{Dassel}, A.~{Eckart} and V.~{Karas}, {Candidate young stellar objects in the S-cluster: Kinematic analysis of a subpopulation of the low-mass G objects close to Sgr A*}, {\em Astronomy and Astrophysics} {\bf 686}, p. A235 (June 2024).

\bibitem{Robitaille2011}
T.~P. {Robitaille}, {HYPERION: an open-source parallelized three-dimensional dust continuum radiative transfer code}, {\em Astronomy and Astrophysics} {\bf 536}, p. A79 (December 2011).

\bibitem{Shahzamanian2016}
B.~{Shahzamanian}, A.~{Eckart}, M.~{Zaja{\v{c}}ek}, M.~{Valencia-S.}, N.~{Sabha}, L.~{Moser}, M.~{Parsa}, F.~{Peissker} and C.~{Straubmeier}, {Polarized near-infrared light of the Dusty S-cluster Object (DSO/G2) at the Galactic center}, {\em Astronomy and Astrophysics} {\bf 593}, p. A131 (October 2016).

\bibitem{Witzel2017}
G.~{Witzel}, B.~N. {Sitarski}, A.~M. {Ghez}, M.~R. {Morris}, A.~{Hees}, T.~{Do}, J.~R. {Lu}, S.~{Naoz}, A.~{Boehle}, G.~{Martinez}, S.~{Chappell}, R.~{Sch{\"o}del}, L.~{Meyer}, S.~{Yelda}, E.~E. {Becklin} and K.~{Matthews}, {The Post-periapsis Evolution of Galactic Center Source G1: The Second Case of a Resolved Tidal Interaction with a Supermassive Black Hole}, {\em Astrophysical Journal} {\bf 847}, p.~80 (September 2017).

\bibitem{Zajacek2017}
M.~{Zaja{\v{c}}ek}, S.~{Britzen}, A.~{Eckart}, B.~{Shahzamanian}, G.~{Busch}, V.~{Karas}, M.~{Parsa}, F.~{Peissker}, M.~{Dov{\v{c}}iak}, M.~{Subroweit}, F.~{Dinnbier} and J.~A. {Zensus}, {Nature of the Galactic centre NIR-excess sources. I. What can we learn from the continuum observations of the DSO/G2 source?}, {\em Astronomy and Astrophysics} {\bf 602}, p. A121 (June 2017).

\end{thebibliography}

\end{document}